\documentclass[journal]{IEEEtran}
\ifCLASSINFOpdf
\else
\fi
\usepackage{nameref,hyperref}
\usepackage{amsmath,amssymb,amsthm} 
\usepackage{array}
\usepackage{multirow}
\usepackage{algorithm}
\usepackage{algorithmicx}
\usepackage[noend]{algpseudocode}
\usepackage{enumitem}
\usepackage{graphicx}
\usepackage{float}
\usepackage{changepage}
\newtheorem{remark}{Remark}
 \usepackage[style=numeric,sorting=none]{biblatex}
\bibliography{citations}

\usepackage{balance}
\AtBeginBibliography{\small}

\begin{document}
\title{Pandemic infection forecasting \\ through compartmental model \\ and learning-based approaches}


\author{\IEEEauthorblockN{Marianna~Karapitta, Andreas~Kasis, Charithea~Stylianides, Kleanthis~Malialis, Panayiotis~Kolios}\\
%\IEEEauthorblockA{\textsuperscript{*}KIOS Research and Innovation Center of Excellence, University of Cyprus, Nicosia, Cyprus}
\thanks{The authors are with the KIOS Research and Innovation Center of Excellence, University of Cyprus, Nicosia, Cyprus (e-mails: karapitta.marianna@ucy.ac.cy; kasis.andreas@ucy.ac.cy;
charitheastylianides@gmail.com; malialis.kleanthis@ucy.ac.cy; kolios.panayiotis@ucy.ac.cy).}
\thanks{This work was supported by the European Union’s Horizon 2020 research and innovation programme under grant agreement No 739551 (KIOS CoE - TEAMING) and from the Republic of Cyprus through the Deputy Ministry of Research, Innovation and Digital Policy. It was also supported by the CIPHIS (Cyprus Innovative Public Health ICT System) project of the NextGenerationEU programme under the Republic of Cyprus Recovery and Resilience Plan under grant agreement C1.1l2.}}% <-this % stops a space

\maketitle

\begin{abstract}
The emergence and spread of deadly pandemics has repeatedly occurred throughout history, causing widespread infections and loss of life. The rapid spread of pandemics have made governments across the world adopt a range of actions, including non-pharmaceutical measures to contain its impact. However, the dynamic nature of pandemics makes selecting intervention strategies challenging. Hence, the development of suitable monitoring and forecasting tools for tracking infected cases is crucial for designing and implementing effective measures. Motivated by this, we present a hybrid pandemic infection forecasting methodology that integrates compartmental model and learning-based approaches. In particular, we develop a compartmental model that includes time-varying infection rates, which are the key parameters that determine the pandemic's evolution. To identify the time-dependent infection rates, we establish a hybrid methodology that combines the developed compartmental model and tools from optimization and neural networks. Specifically, the proposed methodology estimates the infection rates by fitting the model to available data, regarding the COVID-19 pandemic in Cyprus, and then predicting their future values through either a)~extrapolation, or b)~feeding them to neural networks. The developed approach exhibits strong accuracy in predicting infections seven days in advance, achieving low average percentage errors both using the extrapolation (9.90\%) and neural network (5.04\%) approaches.
\end{abstract}

%\begin{IEEEkeywords}
%IEEE, IEEEtran, journal, \LaTeX, paper, template.
%\end{IEEEkeywords}

\IEEEpeerreviewmaketitle

\section{Introduction}
According to the World Health Organization (WHO), a pandemic is a public health emergency of international concern, defined as “an extraordinary event which is determined to constitute a public health risk to other States through the international spread of disease and to potentially require a coordinated international response”. This definition implies a situation that is serious, unexpected, carries implications for public health beyond the affected state's national border, and requires immediate international action~\cite{Q&AWHO}. Over the years there have been numerous pandemics that caused a significant burden on global economies and public health~\cite{qiu2017pandemic}. Existing evidence suggests that the likelihood of pandemics has increased during the past century~\cite{jones2008global, morse2001factors}.

Among a large number of historical pandemics, one that made humanity to highly suffer, killing tens of millions of people, is the Cholera pandemic. The first of seven cholera pandemics emerged in India in 1817 and was caused by the ingestion of food or water contaminated with the bacterium Vibrio cholerae~\cite{WHOcholera}. The Acquired Immunodeficiency Syndrome (AIDS) is a pandemic, caused by the HIV virus and was first identified in 1981. AIDS has claimed 40.1 million lives so far with global ongoing transmission and remains a major global public health issue~\cite{WHOaids}. First identified in 2003, Severe Acute Respiratory Syndrome (SARS) is believed to have possibly originated from bats and spread to humans in China~\cite{WHOsars}. SARS is characterized by respiratory problems, fever and body aches and is spread through respiratory droplets from coughs and sneezes. Quarantine efforts proved effective in controlling the virus, and since then the disease has not reappeared.

At the beginning of December 2019, a “pneumonia of unknown etiology" outbreak was reported in Wuhan, China~\cite{li2020early}. Soon, the cause proved to be a new coronavirus associated with the Severe Acute Respiratory Syndrome virus (SARS-CoV), which later has been termed coronavirus disease 19 (COVID-19)~\cite{coronaviridae2020species}. The rapid spread of the virus to all Chinese provinces and, as of 1st March 2020, to 58 other countries~\cite{li2020substantial}, led the WHO to officially declare it a pandemic. Since then, the world has been facing an unprecedented human crisis with governments across the planet taking intervention measures to curtail the spread of the disease. Some efforts to contain the pandemic included a range of non-pharmaceutical interventions such as closing schools, banning public events, suspension of all public transport, self-isolation, and lockdown policies. Given the large uncertainty regarding the virus transmission, government intervention policies varied considerably among countries. In addition, although such interventions may cope with the spread of the virus~\cite{maharaj2012controlling,maier2020effective}, they impose a vital disruption to the economic and social structure globally. The latter has lead studies to consider the trade-offs between the pandemic effects and the economic impact resulting from the imposed interventions, and proposing suitable strategies \cite{kasis2022optimal, bin2021post}.

Modelling and forecasting the disease's spread is key for the assessment of its transmissibility, and the design of suitable intervention measures \cite{bertozzi2020challenges}. 
Predicting the growth of infections may inform governments and healthcare professionals about an upcoming wave, and enable appropriate intervention strategies.
Thus, accurately predicting infections is critical for effectively managing a pandemic.

Traditionally, compartmental models have proved to be valuable instruments in describing infectious disease epidemics and facilitating the characterization of their asymptotic behaviour and dependence on model parameters. 
Various approaches to model and predict the progression of the aforementioned COVID-19 outbreak rely on these models. Such models enable the study of the disease, within a population divided into compartments, from which they are traditionally named. An extensive review of compartmental models can be found in~\cite{hethcote2000mathematics}. A classic example of a compartmental 50
model is the SIR model~\cite{kermack1927contribution,blackwood2018introduction}. In the SIR model, the population is split into three compartments: Susceptible (S) are the individuals who have no immunity and might become infected if exposed, Infected (I) are the individuals who are currently infected and can transmit the disease to susceptible individuals, and Recovered (R) are the individuals who are immune to the disease and do not transmit the disease to others. The variation of each compartment with time is described by a set of equations with suitable parameters describing the rates of transition among states~\cite{brauer2012mathematical}. Extensions of current models may involve the incorporation of additional compartments, such as Exposed, Diagnosed, Ailing, Threatened, Extinct, and Vaccinated. An eight-compartment model for COVID-19 is proposed in~\cite{giordano2020modelling}, which separates diagnosed and non-diagnosed infected individuals as the former are typically isolated and hence less likely to spread the disease.

For COVID-19, certain parameters within compartmental models represent crucial information that determines the progression of the pandemic. This lead to studies considering compartmental models with time-varying parameters ~\cite{calafiore2020time, li2021toward}.
These models enable the study of the pandemic evolution by evaluating possible influencing factors such as vaccine availability, restrictive measures by governments, environmental temperature, or changes in virus features. For example,~\cite{calafiore2020time} uses a SIRD (Susceptible, Infected, Recovered, Deceased) model with time-varying parameters to capture possible changes of the epidemic behaviour in Italy, due to measures enforced by authorities or modifications of the epidemic characteristics. Furthermore,~\cite{li2021toward} uses a SEIR (Susceptible, Exposed, Infected, Recovered) model with time-dependent infection rate, recovery rate and reproduction number to predict the spread of the COVID-19 pandemic in the United States.

The continuous evolution of dominant variants and the impact of the SARS-CoV-2 virus underscore the necessity for precise modeling of pandemic behavior, including the identification of the most important influencing factors. Predicting infection rates involves analyzing past data, acknowledging their highly variable and dynamic nature. 

Machine learning approaches have been widely employed for this effort, demonstrating notable effectiveness in the case of COVID-19~\cite{malki2020association,dandekar2020quantifying}. Incremental learning approaches \cite{elwell2011incremental, ditzler2015learning, malialis2021online, malialis2022nonstationary}, enable the necessary flexibility to achieve optimized solutions in non-stationary environments. In~\cite{Farooq2020-bi} and~\cite{Farooq2021-hq}, incremental learning of a neural network calculates five parameters (rate of infection during lockdown, time lockdown begins, rate of death, rate of recovery) of a SIRVD (Susceptible, Infected, Recovered, Vaccinated, Deceased) model. 
Moreover, \cite{Farooq2020-bi} aims to forecast the monthly deceased population under different scenarios, while \cite{Farooq2021-hq} focuses on predicting the monthly total number of cases, active infections and deaths.
Furthermore, the hybrid approach developed in~\cite{Camargo2022-kq} uses data on the exposed, infected, recovered and dead population with a combination of compartmental models and a statistical ARIMA model to identify the most suitable time series window that optimizes prediction. 
Then, a data stream and an ensemble of machine learning algorithms incrementally provide predictions at each time step \cite{Camargo2022-kq}. Lastly, there exist works which are solely incremental learning-based  \cite{liu2020real, rodriguez2021deepcovid, stylianides2023study}.

Our study develops a novel compartmental model designed to align with the characteristics of the COVID-19 pandemic. 
The model enables the incorporation of time-varying infection rates, that constitute the key parameters that determine the evolution of the pandemic. In addition, it exhibits strong performance in describing the infected population from our available COVID-19 data regarding the Republic of Cyprus, concerning a population of approximately one million people, yielding a mean absolute percentage error of $3.01\%$. The time-varying rates are forecast by a hybrid approach that combines the proposed compartmental model with machine learning approaches. In particular, our study utilizes the available data to estimate the dynamic parameters of the model. Subsequently, predictions of their future values are obtained through either a)~extrapolation, or b)~the employment of neural networks. The neural networks are trained by incremental learning, allowing the continual adaptation of the time-varying rates to the non-stationary characteristics of the pandemic. The produced infection rates are later input into the compartmental model to enable predictions of the infected population. The developed approach demonstrates high predictive accuracy for future infections seven days in advance, with low average percentage errors observed in both extrapolation (9.90$\%$) and neural network (5.04$\%$) approaches.

The contributions of this work are summarized as follows:
\begin{enumerate}[label = (\roman*)]
    \item We propose a novel compartmental model that involves time-dependent infection rates, which are the parameters with the largest impact on the pandemic progression. The proposed model is able to describe the infected population with high accuracy, yielding a mean average percentage error of 3.01\%.
    
    \item We establish a hybrid approach, that combines a compartmental model with machine learning, suitable for the identification of the time-varying infection rates. The developed approach achieves significantly low errors in predicting future infections seven days in advance,  both through the extrapolation (9.90\%) and neural network (5.04\%) approaches, which are close to the model's ability to describe the data.
\end{enumerate}

\section{Results and Discussion}
\label{ResultsAndDiscussion}

To study the progression of the COVID-19 pandemic, we developed the compartmental SIDAREVH model with time-varying infection rates, which significantly enhance the model's ability to describe the data associated with the pandemic evolution. The estimation of the time-varying infection rate values was achieved through two developed approaches explained in subsections~\nameref{EstimationOfParameters} and \nameref{PredictionOfParametersML}. 
The identified parameter values were then utilized to develop predictions of the infected population, following the approach described in subsection~\nameref{Prediction}. All subsections referenced above are detailed in section~\nameref{Methods}. The proposed approach's forecasting accuracy was evaluated using the mean percentage error between predicted values and data (see subsection \nameref{Prediction}). Below, we present the developed SIDAREVH model, and the main results concerning (i)~the estimation of the time-varying infection rates, associated with the developed SIDAREVH model, following extrapolation and neural network approaches, and (ii)~the accuracy of our approach in predicting future infections.

\subsection{The SIDAREVH compartmental model}
\begin{figure}[!ht]
    \centering
    \includegraphics[scale=0.45]{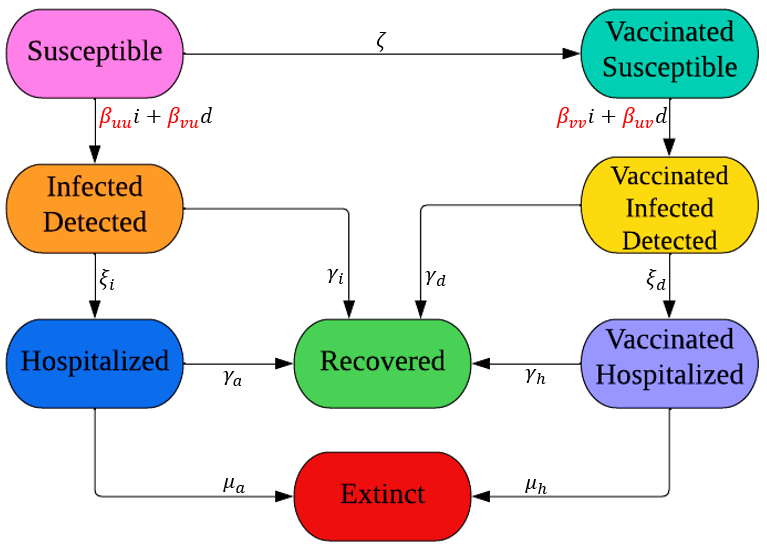}
    \caption{{\bf The SIDAREVH model.}
    Schematic representation of the SIDAREVH model used to describe the evolution of the COVID-19 pandemic. The model splits the population into susceptible, vaccinated susceptible, unvaccinated infected detected, vaccinated infected detected, unvaccinated hospitalized, vaccinated hospitalized, recovered and extinct. Model parameters $\zeta$, $\beta_{uu}$, $\beta_{vu}$, $\beta_{vv}$, $\beta_{uv}$, $\gamma_{i}$, $\gamma_{d}$, $\gamma_{\alpha}$, $\gamma_{h}$, $\xi_{i}$, $\xi_{d}$, $\mu_{\alpha}$, $\mu_{h}$ indicate the transition rates between the states, where $\beta_{uu}$, $\beta_{vu}$, $\beta_{vv}$, and $\beta_{uv}$ are considered as time varying while the remaining as constant.}
\label{fig:SIDAREVHmodel}
\end{figure}

The SIDAREVH compartmental model is a deterministic model developed to characterize the progression of the COVID-19 pandemic. The model was named SIDAREVH, after its compartments, into which it partitions the investigated population. The proposed compartments are: susceptible (S), unvaccinated infected detected (I), vaccinated infected detected (D), unvaccinated hospitalized (A), recovered (R), extinct (E), vaccinated susceptible (V), and vaccinated hospitalized (H). A schematic representation of the SIDAREVH model is shown in Fig~\ref{fig:SIDAREVHmodel}.

The SIDAREVH model includes a large number of parameters that describe the 
transition rates between its compartments. The parameters with the largest impact on the evolution of the pandemic are the infection rates $\beta_{uu}$, $\beta_{vu}$, $\beta_{vv}$, $\beta_{uv}$ which are considered time-varying to account for the effect of factors such as changes in the virus features, control measures, and environmental characteristics. These infection rates (IR) are described in Table \ref{tab:descriptionofIR} below.

\begin{table}[ht!]
\centering
\caption{{\bf Description of the time-varying infection rates involved in the SIDAREVH model.}}
    \begin{tabular}{|c|l|}
    \hline
    \bf Parameter & \bf Description \\ \hline
    $\beta_{uu}$ & Unvaccinated to unvaccinated infection rate\\ \hline
    $\beta_{vu}$ & Vaccinated to unvaccinated infection rate \\ \hline
    $\beta_{vv}$ & Vaccinated to vaccinated infection rate\\ \hline
    $\beta_{uv}$ & Unvaccinated to vaccinated infection rate \\ \hline
    \end{tabular}
\label{tab:descriptionofIR}
\end{table}

A full mathematical description and suitable explanations regarding the validity of the SIDAREVH model are provided in section~\nameref{SIDAREVHmodel}.

\subsection{Infection rates model-based estimation and learning-based prediction}

\subsubsection{Model-based estimated infection rates}
\begin{figure*}[!ht]
\begin{center}
\centering
\includegraphics[width=1\linewidth]{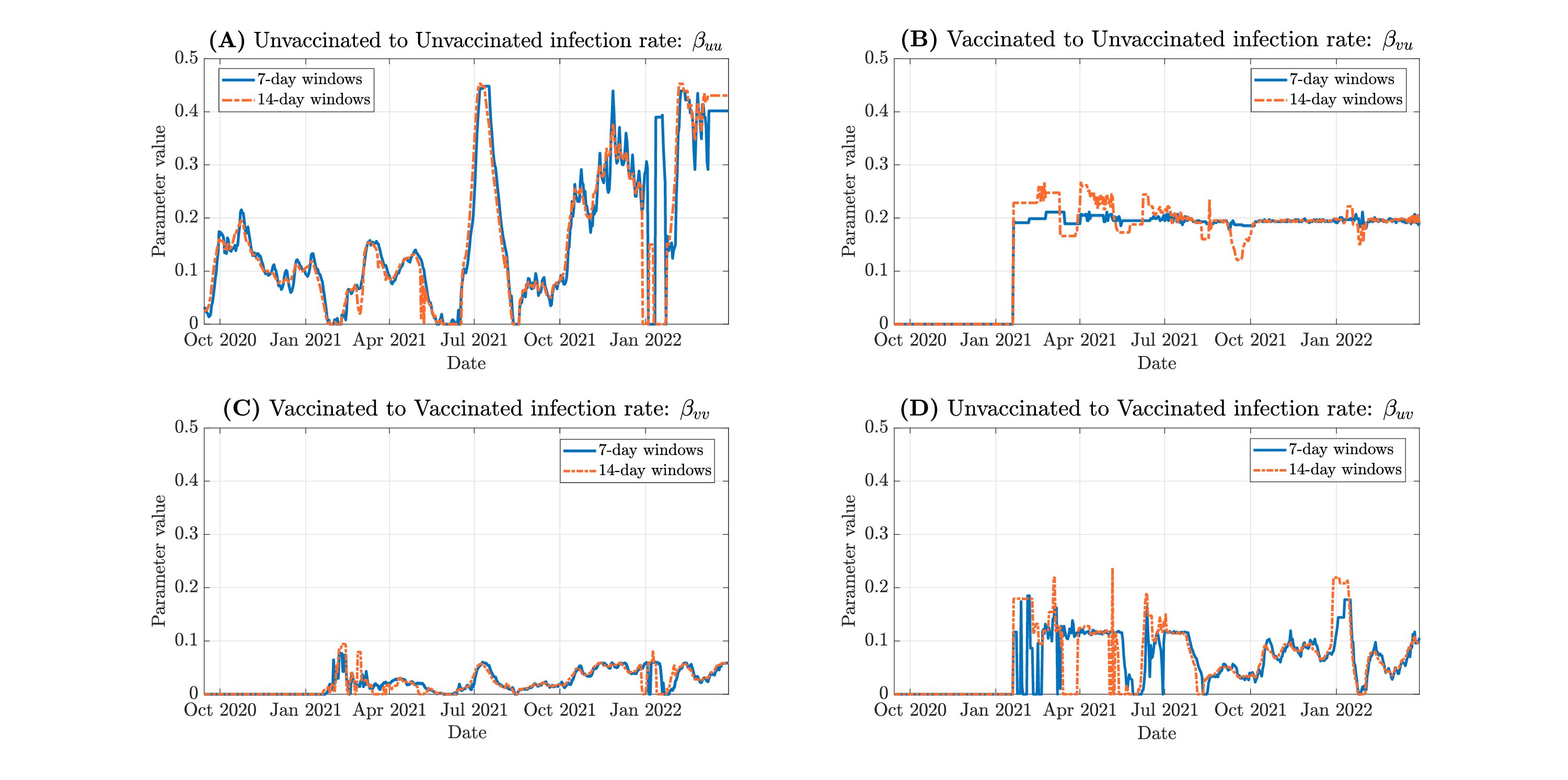}
\end{center}
\caption{{\bf The model-based estimated infection rate (IR) values associated with the SIDAREVH model.} The model-based values of the time-dependent infection rates, using sliding windows of seven and fourteen days: \textbf{(A)} Unvaccinated to Unvaccinated IR: $\beta_{uu}$, \textbf{(B)} Vaccinated to Unvaccinated IR: $\beta_{vu}$, \textbf{(C)} Vaccinated to Vaccinated IR: $\beta_{vv}$, and \textbf{(D)}~Unvaccinated to Vaccinated IR: $\beta_{uv}$.}
\label{fig:estimatedparameters}
\end{figure*}

The results of the estimated time-dependent infection rates $\beta_{uu}, \beta_{vu}, \beta_{vv}, \beta_{uv}$ involved in the SIDAREVH model, using seven and fourteen-day estimation windows are shown in Fig~\ref{fig:estimatedparameters}. It should be noted that the infection rates presented in Fig~\ref{fig:estimatedparameters} were filtered by replacing outliers with the nearest non-outlier value. Details regarding the filtering technique can be found in~\nameref{Appendix}. 

Higher values of infection rates yield a higher frequency of new infections within the population.  The infection rate of the COVID-19 disease is affected by various factors, such as the government's intervention measures~\cite{pan2020association} and the new variants introduced \cite{el2022three} during the examined period. In our considered data, regarding Cyprus, the variants Alpha (B.1.1.7), Delta (B.1.617.2), and Omicron (B.1.1.529) coexisted~\cite{chrysostomou2023unraveling}, being dominant at different periods. In addition, studies such as~\cite{wang2020temperature} and~\cite{tan2005initial} show that transmission rates, and consequently the spread of the SARS-CoV-2 virus, are significantly correlated with the environmental temperature and humidity. 

Fig~\ref{fig:estimatedparameters}A depicts the estimated rate at which unvaccinated people infect other unvaccinated people. This rate takes the highest values among the four rates in Fig~\ref{fig:estimatedparameters}, due to the susceptibility of the unvaccinated individuals to the disease. The $\beta_{uu}$ rates using seven and fourteen-day windows follow the same shape in the entire investigated time, except the period around January 2022 when the Omicron variant became dominant and replaced the Delta variant. In this period, the resulting rate using fourteen-day windows was changing slower and the filtering technique used to identify and fill the outliers decreased these values. Fig~\ref{fig:estimatedparameters}B depicts the estimated rate at which vaccinated people infect unvaccinated people. The estimated parameters indicate that, with both seven and fourteen-day windows, the rates remain nearly constant throughout the entire time period. The rate $\beta_{vv}$ at which vaccinated people infect other vaccinated people, presented in Fig~\ref{fig:estimatedparameters}C, takes the lowest values of all four infection rates, demonstrating the lower infectivity among immunized population. Finally, the rate at which unvaccinated people infect vaccinated people, denoted by $\beta_{uv}$ and presented in Fig~\ref{fig:estimatedparameters}D, also follows the same shape for the entire time period. The three rates presented in Figs~\ref{fig:estimatedparameters}B, C, D, that are associated with vaccinations, are zero until the beginning of 2021 when the vaccinations started in Cyprus~\cite{OurWorldInData}.

The following results demonstrate the model's ability to accurately describe the data. In particular, Table~\ref{tab:seven-day_unfiltered_betas} presents the model's accuracy in terms of describing the infected population by considering seven-day prediction windows. For Table~\ref{tab:seven-day_unfiltered_betas}, the unfiltered estimated infection rates ($\beta_{uu}, \beta_{uv}, \beta_{vu}, \beta_{vv}$) were used (these are the output values from Algorithm~\ref{alg:estimatingmethod} presented in section \nameref{EstimationOfParameters}). These estimated infection rate values, obtained by fitting the model to the data, give a lower bound to the prediction accuracy that the proposed SIDAREVH model may achieve through the use of seven-day windows. As Table~\ref{tab:seven-day_unfiltered_betas} shows, the SIDAREVH model may describe our data with a mean absolute percentage error (MAPE) of 3.01\%. In addition, when the largest 5\% of the absolute errors is excluded then the corresponding error drops to 2.63\%. This 5\% of highest errors largely coincides with periods of abrupt changes in the pandemic behaviour attributed to external factors, and hence not being predictable through the available data, such as when a new variant became dominant or under sudden changes in the intensity of government intervention measures. Excluding the largest 5\% of the errors yields a notable reduction in the MAPE (3.01\% to 2.63\%) and a substantial reduction in the associated standard deviation (S.D.) (2.30\% to 1.44\%). The corresponding results concerning windows of fourteen days are presented in~\nameref{Appendix} and demonstrate higher error values. This lead us to choose seven-day windows, rather than fourteen-day windows, for the estimation of the infection rates.

\begin{table}[!ht]
\centering
\caption{{\bf SIDAREVH model's accuracy in describing the infected population data using model-based optimized unfiltered time-varying infection rates with seven-day windows.}}
    \begin{tabular}{|c|c|c|}
        \hline
        \bf & \bf MAPE & \bf S.D.\\ \hline
        All data & 3.01\% & 2.30\%  \\ \hline
        Lowest 95\% of errors & 2.63\% & 1.44\% \\
        \hline
    \end{tabular}
\label{tab:seven-day_unfiltered_betas}
\end{table}

\subsubsection{Learning-based predicted infection rates}
In this section, we present the results associated with the infection rates prediction using the approach described in section \nameref{PredictionOfParametersML}. The prediction accuracy is evaluated through the widely used regression forecast metrics Mean Absolute Error (MAE) and Mean Absolute Percentage Error (MAPE) (see Eqs (\ref{eq:mae}) and (\ref{eq:mape} in \nameref{Methods}). Furthermore, all experiments are run over 10 repetitions, with a Multilayer Perceptron approach, and the mean and standard deviation (S.D.) of the relevant metrics are presented. 

We have examined the performance of the proposed method using three lookback window sizes $W \in \{7, 14, 30\}$ (see section \nameref{PredictionOfParametersML}). In all cases, the four considered infection rates were forecast a week ahead. The best performing lookback window size, presented in Table \ref{tab:nnperformance} below is with a fourteen-day window. It should be clarified that the windows for the infection rates estimation, which was selected to be seven-days, is different than the lookback window, which is fourteen-days. The latter determines how many past values of infection rate estimates are considered for prediction. For comparison, the results for all window sizes are shown in \nameref{Appendix}.

\begin{table*}[h!]
\centering
\caption{\bf Multilayer Perceptron performance for each infection rate, for a lookback window size of fourteen days, in terms of mean absolute error, mean absolute percentage error and associated standard deviations.}
\label{tab:nnperformance}
    \begin{tabular}{|c|c|c|c|c|}
    \hline
     & $\beta_{uu}$ & $\beta_{vu}$ & $\beta_{vv}$ & $\beta_{uv}$ \\ \hline 
     \textbf{MAE (S.D.)} & 0.011 (0.015) & 0.003 (0.008) & 0.003 (0.005) & 0.006 (0.012) \\ \hline
     \textbf{MAPE (S.D.)} & 6.346\% (7.567\%) & 1.483\% (4.047\%) & 15.403\% (17.779\%) & 7.996\% (13.105\%) \\ \hline
    \end{tabular}
    \begin{flushleft}
        The infection rates are described as $\beta_{uu}$: Unvaccinated $\rightarrow$ Unvaccinated IR, $\beta_{vu}$: Vaccinated $\rightarrow$ Unvaccinated IR, $\beta_{vv}$: Vaccinated $\rightarrow$ Vaccinated IR, $\beta_{uv}$: Unvaccinated $\rightarrow$ Vaccinated IR.
    \end{flushleft}
\end{table*}

\begin{figure*}[!ht]
        \centering
        \includegraphics[width=1\linewidth]{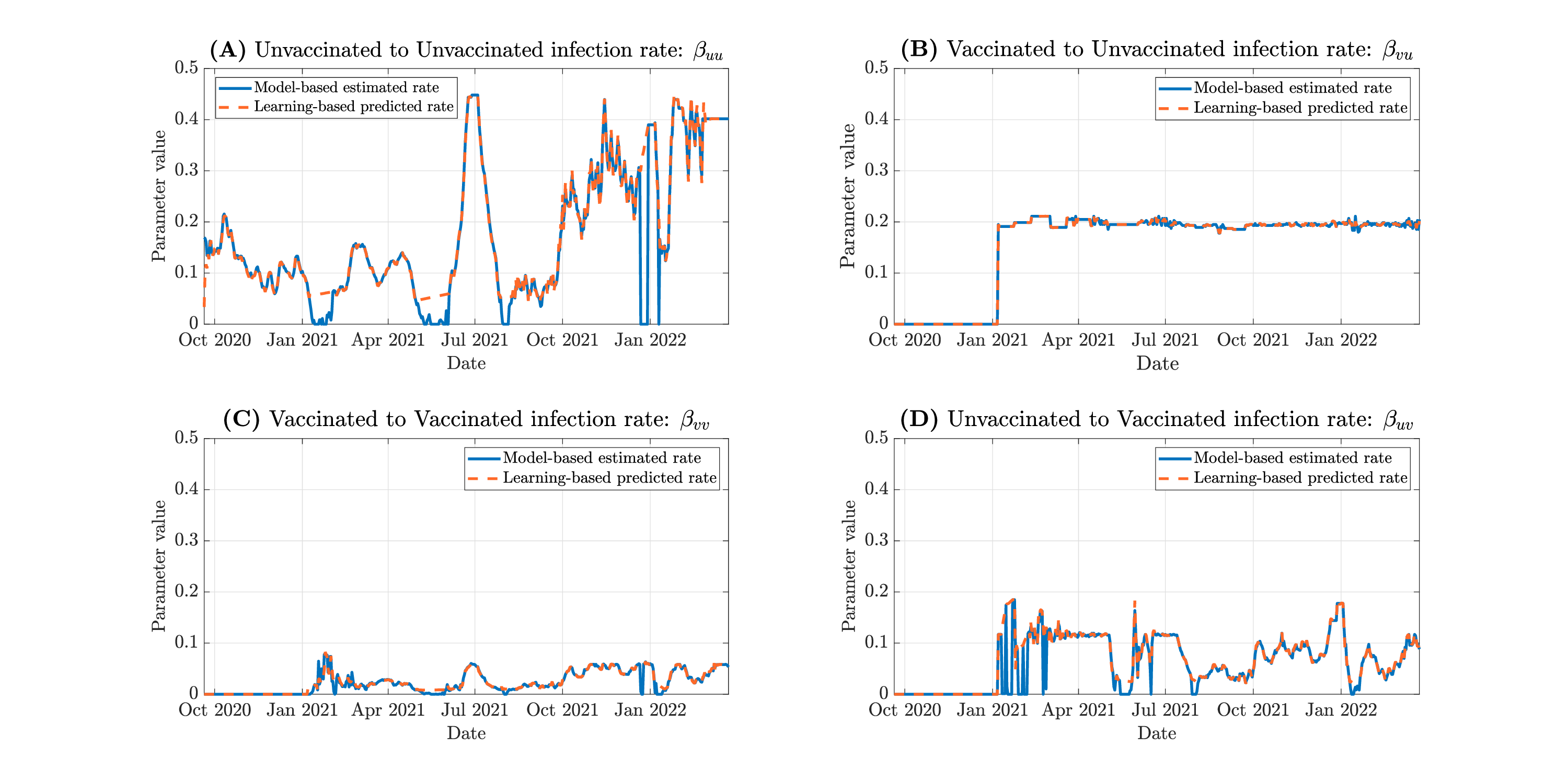}
        \caption{{\bf  Model-based estimated versus learning-based predicted infection rate (IR) values associated with the SIDAREVH model.} \textbf{(A)} Unvaccinated to Unvaccinated IR: $\beta_{uu}$, \textbf{(B)} Vaccinated to Unvaccinated IR: $\beta_{vu}$, \textbf{(C)} Vaccinated to Vaccinated IR: $\beta_{vv}$, and \textbf{(D)} Unvaccinated to Vaccinated IR: $\beta_{uv}$.}
        \label{fig:predictedparameters}
\end{figure*}

In addition to Table \ref{tab:nnperformance}, the performance of the predicted infection rates compared to the estimated ones is depicted in Fig~\ref{fig:predictedparameters}, which demonstrates that the estimated and forecasted values are very close, despite the dynamic nature of the pandemic. Important to be noted is that the infection rate values are directly affected by policy changes and weather-related conditions that are not reflected in the data, and hence some error margin in their forecast is expected, regardless of the adopted approach.

\subsection{Model and learning-based prediction of infected population}

\subsubsection{Model-based prediction of infected population}
The prediction results presented below were generated using the estimated infection rates obtained from Algorithm~\ref{alg:estimatingmethod} (see section \nameref{EstimationOfParameters}). The projections refer to the infected population (both vaccinated and unvaccinated). Table~\ref{tab:predictionstatistics_first} presents the model's seven-day predictive accuracy, in terms of mean absolute percentage errors, utilizing the filtered estimated infection rates, illustrated in Fig~\ref{fig:estimatedparameters}, where the outlier values were removed. 
This approach extrapolates the optimized estimated infection rates, obtained at each day using the previous seven-day window, to the next seven-day window, as explained in section \nameref{Prediction}. Essentially, using past week's information, this approach enables predictions for the infected population for the subsequent seven days. As shown in Table \ref{tab:predictionstatistics_first}, this results in a mean absolute  percentage error of 9.90\% (standard deviation 8.05\%), which decreased to 8.70\% (standard deviation 6.22\%) when the highest 5\% of errors was excluded.

\begin{table}[!ht]
\centering
\caption{{\bf SIDAREVH model's accuracy in describing the infected population data using model-based optimized filtered time-varying infection rates with seven-day windows.}}
\begin{tabular}{|c|c|c|}
\hline
 & \bf MAPE & \bf S.D.\\ \hline
All data & 9.90\% & 8.05\%  \\ \hline
Lowest 95\% of errors & 8.70\% & 6.22\% \\ \hline
\end{tabular}
\label{tab:predictionstatistics_first}
\end{table}

Predictions of the infected population were also generated by utilizing fourteen-day windows of previous data, through a similar approach. The corresponding predictive performance and errors are available in \nameref{Appendix}. We chose to only present the prediction errors using seven-day windows of previous data since, as demonstrated in \nameref{Appendix}, the average daily error of the predicted windows increases when fourteen-day windows of previous data are considered, yielding a higher average percentage prediction error. This holds regardless of the prediction timeframe, i.e. regardless of how many days in the future we forecast.  

\subsubsection{Learning-based prediction of infected population}
The prediction results presented below were generated using the infection rates obtained from Algorithm~\ref{alg:nnmethod} (see section \nameref{PredictionOfParametersML}). The forecasts correspond to seven-day sliding prediction windows utilizing the infection rates that best describe the preceding seven-day period. Table~\ref{tab:predictionstatistics_ML_first} presents the error statistics of the predictions created using the infection rates resulting from the neural network approach presented in section \nameref{PredictionOfParametersML}. This approach enables a mean absolute percentage error of 5.04\% (standard deviation 7.03\%), which drops to 3.74\% (standard deviation 4.27\%) when the highest 5\% error values are excluded. Hence, this approach enables a significant improvement in terms of forecasting accuracy, compared to the results presented in Table~\ref{tab:predictionstatistics_first}. They also demonstrate the strong performance of the developed approach, since the results are close to those depicted in Table~\ref{tab:seven-day_unfiltered_betas}, concerning the SIDAREVH's model ability to describe the data, that can be interpreted as a lower bound to the errors that may be obtained using this model.

\begin{table}[!ht]
\centering
\caption{{\bf SIDAREVH model's accuracy in describing the infected population data using learning-based predicted time-varying infection rates with seven-day windows.}}
\begin{tabular}{|c|c|c|}
\hline
\bf   & \bf MAPE & \bf S.D.\\ \hline
All data & 5.04\% & 7.03\%  \\ \hline
Lowest 95\% of errors & 3.74\% & 4.27\% \\
\hline
\end{tabular}
\label{tab:predictionstatistics_ML_first}
\vspace{-3mm}
\end{table}

\subsubsection{Combined forecasts}
\begin{figure}[!ht]
\centering
\includegraphics[scale=0.6]{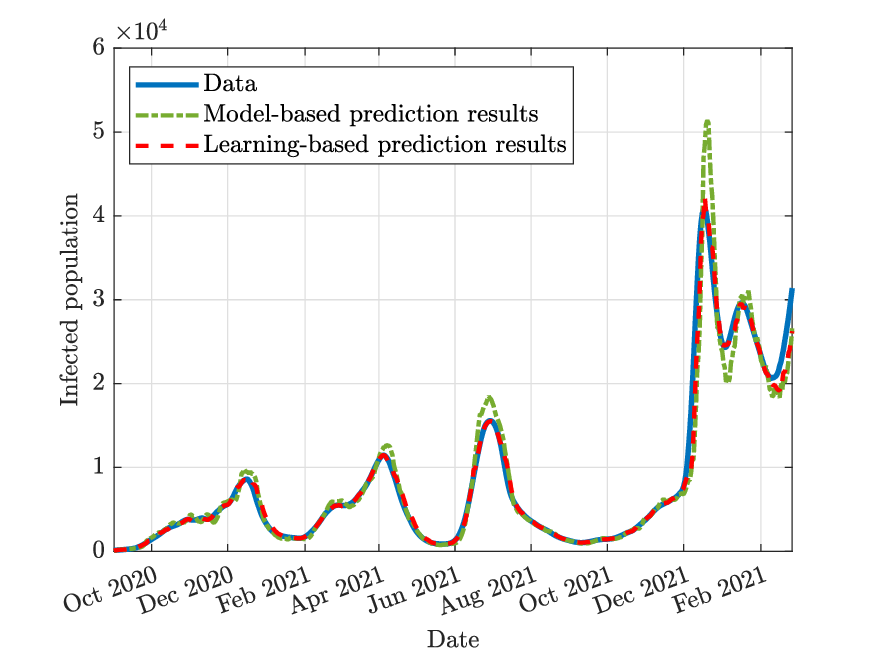}
\caption{{\bf Prediction results.} Schematic representation of the model and learning based prediction results compared to the data. The daily values of prediction refer to the average infected population, considering seven-day windows.}
\label{fig:predictionresults_first}
\end{figure}

A graphical representation of the predicted infected population during the examined period, is depicted in Fig~\ref{fig:predictionresults_first}. 
In particular, Fig~\ref{fig:predictionresults_first} illustrates the average predicted infected population of every seven-day prediction sliding window, as generated by using the infection rates identified by the two proposed methodologies, in comparison with the real data. The figure shows that, in both cases, the predictions are very close to the data, demonstrating the performance of the established approaches and the suitability of the developed SIDAREVH model.

\section{Materials and methods}
\label{Mataerials_and_methods}
\begin{figure}[h!]
\centering
        \centering
        \includegraphics[scale=0.35]{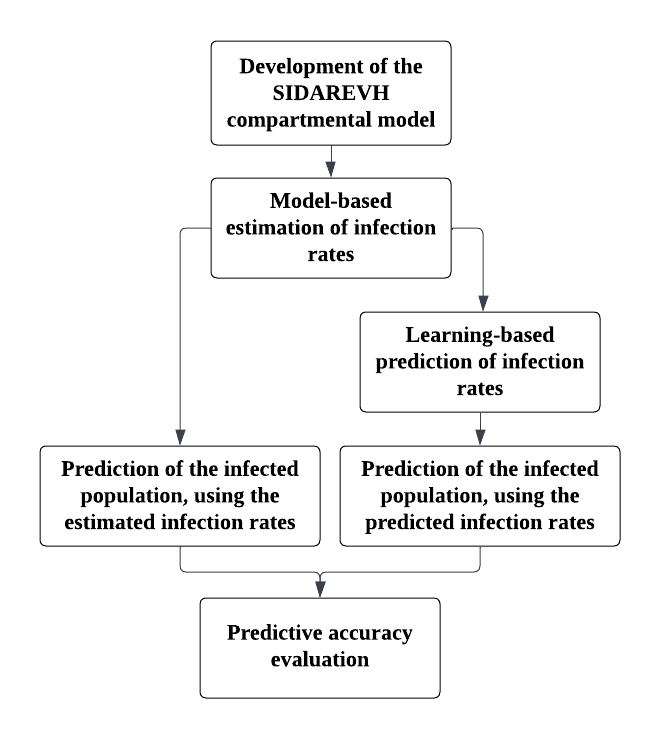}
        \caption{\textbf{Flowchart of the steps followed in the proposed methodology.}}
        \label{fig:Process}
\end{figure}

This section describes the available data regarding the COVID-19 pandemic in the Republic of Cyprus and the methodologies employed to study and forecast its evolution. Within the \nameref{Data} subsection, we present the available data and explain how they are processed to enable their analysis. The \nameref{Methods} subsection includes five stages that are interdependent. Firstly, in subsection \nameref{SIDAREVHmodel} we present the developed SIDAREVH model and its parameters.
Then, in subsection \nameref{EstimationOfParameters} we describe the model-based approach developed to estimate the time-dependent infection rates included in the SIDAREVH model, and in subsection \nameref{PredictionOfParametersML} we present the learning-based approach to predict the future values of the time-dependent infection rates, based on the model-based approach's results. The parameter values resulting from the two approaches were used for predicting the infected population in short-time windows and our method to evaluate the predictive ability of the developed approaches is explained in \nameref{Prediction}. A graphical representation of the methodology steps is provided in Fig~\ref{fig:Process}. To enhance the readability of the manuscript, the main symbols used through this section are provided in \nameref{Appendix}.

\subsection{Data}
\label{Data}
The data used in this study were provided by the Cyprus Ministry of Health. 
They include anonymous daily COVID-19 confirmed infections and details on their progression for Cyprus, an EU country with a population of $920000$~\cite{Census2021}. Particularly, for every infection case the available data include its confirmation date, vaccination status, hospitalization and discharge dates (when applicable), and recovery or disease dates. Additionally, data about the daily vaccinations in Cyprus were obtained from~\cite{OurWorldInData}. The investigated time period of the collected data was from March 2020 to March 2022.

To best represent the entire population of the available data, we considered the following eight groups, such that each individual falls in one of them. Moreover, these groups are associated to the states of the SIDAREVH compartmental model described above. The eight groups are given as follows: susceptible, unvaccinated infected, unvaccinated hospitalized, vaccinated infected, vaccinated hospitalized, recovered, and deceased. These groups are sufficient to describe the status of each individual in relation to the pandemic at each time. For each day, we calculated the number of individuals that belonged to each of the eight considered groups. Additionally, we considered that initially, the total population was susceptible. The remaining groups were initialized at zero and modified based on the confirmed case progression. Specifically, upon infection, susceptible individuals were removed from the susceptible group and added to the unvaccinated infected group. If unvaccinated infected individuals required hospitalization, they were removed from the unvaccinated infected group and added to the unvaccinated hospitalized group; if not, they were removed from the unvaccinated infected group and added to the recovered group. Unvaccinated hospitalized individuals were removed from the unvaccinated hospitalized group and  added either to the recovered or the deceased group. Susceptible individuals who got vaccinated were removed from the susceptible and added to the vaccinated group. Then, the same group transitions pattern for unvaccinated individuals was followed for the vaccinated individuals.

\begin{figure*}[h!]
\centering
    \includegraphics[width=1\linewidth]{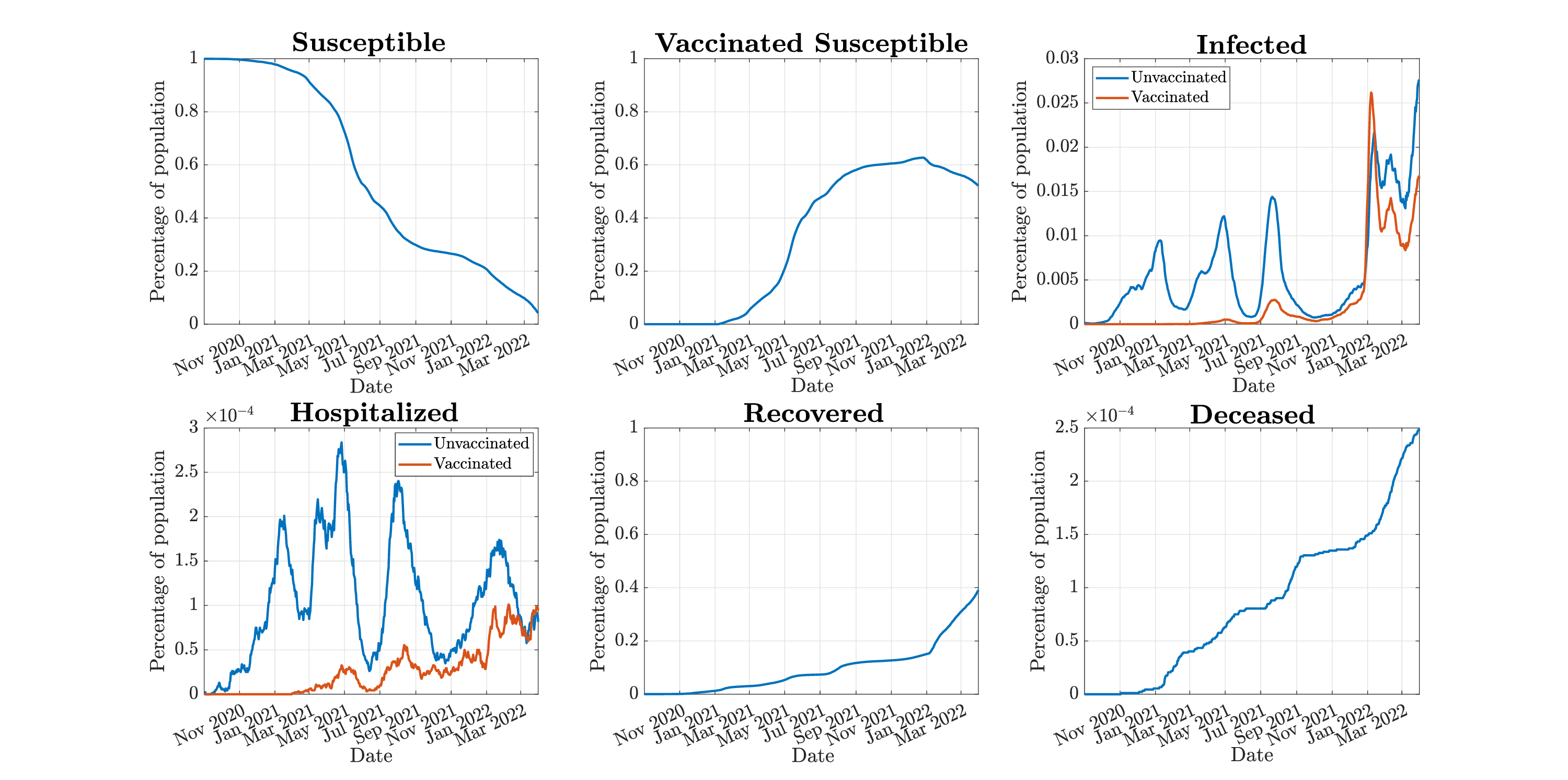}
    \caption{{\bf Groups' progression over time, as calculated from the available data.} Percentage of individuals within the susceptible, unvaccinated infected, vaccinated infected, unvaccinated hospitalized, vaccinated hospitalized, recovered and deceased groups of population, from September 2020 to March 2022.}
    \label{fig:trajectories}
\end{figure*}

In Fig~\ref{fig:trajectories} the percentage of each group of the population is graphically demonstrated.
The presented group values refer to the time period from September 2020 to March 2022, which was the period used for the procedures explained in the following sections. Although our dataset includes data from March 2020, the number of cases between March and September 2020 was negligible and therefore not used for analysis purposes.

\subsection{Methods}
\label{Methods}

\subsubsection{Description of the SIDAREVH compartmental model}
\label{SIDAREVHmodel}
The developed SIDAREVH model characterizes the progression of the COVID-19 pandemic and includes eight compartments. The proposed compartments were selected to include all the available data about Cyprus. The dynamics of the SIDAREVH model are given below.

\begin{subequations}
    \begin{eqnarray}
        \dot{s}=-\beta_{uu} is-\beta_{vu} ds-\zeta s,
    \end{eqnarray}
    \begin{eqnarray}
        \dot{i}=\beta_{uu} is + \beta_{vu} ds -\xi_{i}i-\gamma_{i}i,
    \end{eqnarray}
    \begin{eqnarray}
        \dot{d}=\beta_{vv}iv + \beta_{uv}dv-\xi_{d}d - \gamma_{d}d,
    \end{eqnarray}
    \begin{eqnarray}
        \dot{a}=\xi_{i}i-\gamma_{\alpha}\alpha-\mu_{\alpha}\alpha,
    \end{eqnarray}
    \begin{eqnarray}
        \dot{r}=\gamma_{i}i+\gamma_{d}d+\gamma_{\alpha}\alpha+\gamma_{h}h,
    \end{eqnarray}
    \begin{eqnarray}
        \dot{e}=\mu_{\alpha}\alpha+\mu_{h}h,
    \end{eqnarray}
    \begin{eqnarray}
        \dot{v}=\zeta s -\beta_{vv}iv-\beta_{uv}dv,
    \end{eqnarray}
    \begin{eqnarray}
        \dot{h}=\xi_{d}d-\gamma_{h}h-\mu_{h}h,
    \end{eqnarray}
    \begin{eqnarray}
s(0)=s_{0},i(0)=i_{0},d(0)=d_{0},\alpha(0)=\alpha_{0},\\r(0)=r_{0},e(0)=e_{0},v(0)=v_{0},h(0)=h_{0},\nonumber
    \end{eqnarray}
    \label{eq:dynamics}
\end{subequations}
where $s,i,d,\alpha,r,e,v,h \in [0,1]$ are the states of the system, describing the population percentage of susceptible, unvaccinated infected detected, vaccinated infected detected, unvaccinated hospitalized, recovered, extinct, vaccinated susceptible, and vaccinated hospitalized, respectively. The initial values for $s,i,d,\alpha,r,e,v,h$ are denoted by $s_{0},i_{0},d_{0},\alpha_{0},r_{0},e_{0},v_{0},h_{0} \in [0,1]$ respectively. 

The SIDAREVH model parameters are described in Table~\ref{tab:descriptionparameters}. We assume that all parameter values are known and constant, except the four infection rates $\beta_{uu}$, $\beta_{vu}$, $\beta_{vv}$, and $\beta_{uv}$. These are the key parameters that govern the pandemic dynamics, since they determine the effect of the included bilinear terms in Eq~(\ref{eq:dynamics}), and hence have the largest impact on the progression of the COVID-19 pandemic. Due to their association with the non-linear dynamics and the effect of influencing factors on them, such as government intervention measures and changing dominant variants, they are assumed to be time-varying. Estimations of the infection rates {and predictions on their future values} are made using the approaches described in sections~\nameref{EstimationOfParameters} and~\nameref{PredictionOfParametersML} respectively. The constant parameters involved in the SIDAREVH model have a smaller impact on the pandemic infections. They were obtained from existing literature studies and are illustrated in~\nameref{Appendix}.

\begin{table*}[ht!]
\centering
\caption{{\bf Description of the parameters involved in the SIDAREVH model.}}
\begin{tabular}{|c|l|}
\hline
\bf Parameter & \bf Description \\ \hline 
$\zeta$ & Rate of vaccination of susceptible individuals\\ \hline
$\beta_{uu}$ & Time-dependent infection rate at which unvaccinated people infect other unvaccinated people \\ \hline
$\beta_{vu}$ & Time-dependent infection rate at which vaccinated people infect unvaccinated people \\ \hline
$\beta_{vv}$ & Time-dependent infection rate at which vaccinated people infect other vaccinated people \\ \hline
$\beta_{uv}$ & Time-dependent infection rate at which unvaccinated people infect vaccinated people \\ \hline
$\gamma_{i}$ & Recovery rate for unvaccinated infected detected individuals \\ \hline
$\gamma_{d}$ & Recovery rate for vaccinated infected detected individuals \\ \hline
$\gamma_{\alpha}$ & Recovery rate for unvaccinated hospitalized individuals \\ \hline
$\gamma_{h}$ & Recovery rate for vaccinated hospitalized individuals \\ \hline
$\xi_{i}$ & Rate at which unvaccinated infected detected individuals become hospitalized  \\ \hline
$\xi_{d}$ & Rate at which vaccinated infected detected individuals become hospitalized\\ \hline
$\mu_{\alpha}$ & Rate at which unvaccinated hospitalized individuals decease \\ \hline 
$\mu_{h}$ & Rate at which vaccinated hospitalized individuals decease \\ \hline
\end{tabular}
\label{tab:descriptionparameters}
\end{table*}

The SIDAREVH model is based on the following assumptions: \newline
\emph{(i)} The considered population is constant, i.e. births and deaths not attributed to COVID-19 are not considered. \newline
\emph{(ii)} The considered population is isolated and imported cases are not included. \newline
\emph{(iii)} Infected individuals become first hospitalized before they decease. \newline
\emph{(iv)} Vaccinations are performed on susceptible individuals only.\newline
\emph{(v)}  Recovered individuals are no longer susceptible to the disease because of acquired immunity.\newline
\emph{(vi)} Vaccinated individuals cannot be susceptible again.\newline
The constant population assumption suggests that the model states satisfy at all times the condition $s+i+d+a+r+e+v+h=1$. The above assumptions constitute theoretical requirements for compartmental models.
\begin{remark}
     It is important to note that, although some of these assumptions may be only partly satisfied in practice, this does not hinder the applicability of our SIDAREVH model. The latter is justified by Table~\ref{tab:seven-day_unfiltered_betas} in section \nameref{ResultsAndDiscussion}, which shows that the SIDAREVH compartmental model is able to describe the infected population with an average percentage error of $3.01\%$. The suitability of the considered assumptions also follows from the fact that this study aims to obtain predictions on a relatively short time horizon, using the available data at the time of the prediction. Although relaxing some of the above assumptions, e.g. by introducing re-susceptibility of the recovered population, could enable a better characterization of the overall pandemic progression \cite{batistela2021sirsi}, this would only have a negligible effect on the forecasting accuracy on such short time windows. 
\end{remark}

\subsubsection{Model-based infection rates estimation}
\label{EstimationOfParameters}
This section describes the methodology employed for the estimation of the infection rates $\beta_{uu}$, $\beta_{vu}$, $\beta_{vv}$, $\beta_{uv}$ associated with the proposed SIDAREVH model. These four parameters are associated with the bilinear terms within the dynamics of Eq~(\ref{eq:dynamics}), and constitute the parameters with the largest impact on the evolution of the infected cases in the pandemic. They are assumed to be time-varying since they are affected by different factors, such as the government intervention measures, and environmental and virus features. Estimates of the infection rates were produced by fitting the SIDAREVH model to the observed time series, depicted in Fig~\ref{fig:trajectories}. The procedure was made through sliding windows of seven or fourteen days, which moved by one day per increment, between September 2020 and March 2022.

To estimate the parameter values, we aimed to minimize the optimization cost function:
\begin{eqnarray}
\label{eq:costfunction}
	 C_{j}(n,i,d,\hat{i}_{j},\hat{d}_{j})=\sum_{t=j+1}^{j+n} [(i(t)-\hat{i}_{j}(t))^2 +\\(d(t)-\hat{d}_{j}(t))^2]\nonumber,
\end{eqnarray}
where $C_{j}$ describes the deviations between the actual and predicted infected cases at day $j$, for a time window of size $n$, where $n\in\{7,14\}$. {Moreover, $i(t)$ and $d(t)$ denote the unvaccinated infected and vaccinated infected states from the data at time $t$ respectively, and $\hat{i}_{j}(t)$ and $\hat{d}_{j}(t)$ are the estimated unvaccinated infected and predicted vaccinated infected states for time $t$ using the knowledge available at time $j$ respectively.}

Algorithm~\ref{alg:estimatingmethod} describes the procedure for estimating the four infection rates $\beta_{uu}$, $\beta_{vu}$, $\beta_{vv}$, $\beta_{uv}$. The aim of this algorithm is to obtain the time-dependent parameter values that minimize the cost defined by Eq~(\ref{eq:costfunction}). The targeted infection rates determine the estimated values $\hat{i}_{j}$ and $\hat{d}_{j}$, and are optimized with aim to minimize the  associated cost, given by Eq~(\ref{eq:costfunction}). Starting from selecting the examined window $z_{j}=\{j-n+1,...,j\}$, initial values for the four infection rates were given, and an initial cost $C_{j}(n,i,d,\hat{i}_{j},\hat{d}_{j})$ was calculated. Subsequently, each of the four rates was increased and decreased by a constant percentage defining the trial parameters set $P_{k}$, that contains the total eight modified rates, where $k$ is the number of iterations. A new cost was calculated for each set of rates and then, the minimum cost $C_{j}'(n,i,d,\hat{i}_{j},\hat{d}_{j})$ of the total eight costs was found. If the minimum cost resulted in a decreased cost compared to the previous iteration, the considered modified infection rate was kept and used for the subsequent iterations, creating the new set of trial parameters $P'_{k}$, until the set of parameters that minimizes the cost of the window $z_{j}$ was found. The same procedure was applied for all the sliding windows $z_{j}$ of the examined period. Finally, four sets of optimal infection rates were obtained, that contained the parameters' time-dependent values that best describe each considered $n$-day window. These four sets are denoted by $\vec{\beta}_{uu}, \vec{\beta}_{vu}, \vec{\beta}_{vv}$, and $\vec{\beta}_{uv}$,  with their values corresponding to the optimal time-depended infection rates $\beta_{uu}, \beta_{vu}, \beta_{vv}$, and $\beta_{uv}$ respectively for all times.

\begin{algorithm}[h!]
	\caption{Optimal infection rates estimation}
	\label{alg:estimatingmethod}
	\begin{algorithmic}[1]
            \Statex \textbf{Inputs:} $s,i,d,a,r,e,v,h$, daily vaccinations
            \Statex \textbf{Output:} 
            {$\vec{\beta}_{uu}, \vec{\beta}_{vu}, \vec{\beta}_{vv}$, $\vec{\beta}_{uv}$} 
            \Statex \textbf{Initialization:} $n=7$ or $n= 14$
            \For{each time step $j$} 
                \State Select window $z_{j}=\{j-n+1,...,j\}$
                \State Initialize parameters $\beta_{uu}$, $\beta_{vu}$, $\beta_{vv}$, $\beta_{uv}$
                \State Calculate the initial cost $C_{j}(n,i,d,\hat{i}_{j},\hat{d}_{j})$
                \State $flag=0$
                \While{$flag=0$}
                    \State Create the trial parameters set $P_{k}$
                    \State Calculate a cost using every parameter from $P_{k}$ set
                    \State Find the minimum of all costs $C_{j}'(n,i,d,\hat{i}_{j},\hat{d}_{j})$ \If{$C_{j}'(n,i,d,\hat{i}_{j},\hat{d}_{j}) < C_{j}(n,i,d,\hat{i}_{j},\hat{d}_{j})$}
                        \State $C_{j}(n,i,d,\hat{i}_{j},\hat{d}_{j}) = C_{j}'(n,i,d,\hat{i}_{j},\hat{d}_{j})$
                        \State Keep the modified infection rate and create new set of trial parameters $P_{k}'$
                        \Else
                        \State $flag=1$
                        \State {$\vec{\beta}_{uu}(j)=\beta_{uu}$}
                        \State{$\vec{\beta}_{vu}(j) =\beta_{vu}$}
                        \State{$\vec{\beta}_{vv}(j)=\beta_{vv}$}
                        \State{$\vec{\beta}_{uv}(j)=\beta_{uv}$}
                    \EndIf
                \EndWhile
            \EndFor
	\end{algorithmic}
\end{algorithm}

\subsubsection{Learning-based infection rates prediction and evaluation} 
\label{PredictionOfParametersML}
We consider the set of the four positive infection rates sets $\mathcal{B} = \{\vec{\beta}_{uu}, \vec{\beta}_{vu}, \vec{\beta}_{vv}, \vec{\beta}_{uv}\}$ as generated by Algorithm \ref{alg:estimatingmethod}. Without any loss of generality, let $\vec{\beta} \in \mathcal{B}$ be any of the aforementioned infection rate sets. Each $\vec{\beta} \in \mathbb{R}^T,  T \in [1, \infty)$  constitutes a univariate time series, defined as $\vec{\beta} = \{\beta^j\}^T_{j=1}$, where {$j$ is the current day,} and $\beta^j$ is the infection rate at day $j$. To address the temporal correlations encountered in the time-series data, we use a sliding or ``lookback'' window, such that, $\grave{\beta}^j = \{\beta^j, \beta^{j-1}, ..., \beta^{j-W+1}\} \in \mathbb{R}^W$ where $W \in \mathbb{N}$ is the window size. Our proposed method was evaluated across three window sizes ($W \in \{7, 14, 30\}$), with the optimal performance consistently observed with a fourteen-day window. For comparison, the results for all window sizes are shown in \nameref{Appendix}.

Let $\mathcal{F} = \{f_{uu}, f_{vu}, f_{vv}, f_{uv}\}$ be a set of four regression models, each corresponding to a time series $\vec{\beta} \in \mathbb{R}^T$ used for forecasting.
{At each day $j \geq W + n - 1$, the aim is to forecast the corresponding infection rate $D$ days later,  by using $\grave{\beta}^j$.}
Without any loss of generality, let $f_\theta \in \mathbb{F}$ be a regression model (e.g., a neural network) parameterised by ${\theta}$, defined as $f_\theta: \mathbb{R^W} \rightarrow  \mathbb{R}$, such that, $\hat{\beta}^{j+D} = f_\theta(\grave{\beta}^j)$. 

At any day $j \geq W + D$, the loss function used between a prediction $\hat{\beta}^j = f_\theta(\grave{\beta}^{j-D})$ and ground truth $\beta^j$ is the Mean Squared Error (MSE) defined as:
\begin{equation}\label{eq:mse}
    Q^j = (\hat{\beta}^j - \beta^j)^2.
\end{equation}

The model gradually adapts without complete re-training using incremental learning, that is, $f_\theta^j = f_\theta^{j-1}.train(\grave{\beta}^{j-D}, \beta^{j})$. It is continually updated using the online Stochastic Gradient Descent approach (or any Gradient Descent-based algorithm) where each model parameter $\theta_i \in \theta$ is updated according to the following formula
\begin{equation}\label{eq:training}
    \theta^j_i \leftarrow \theta^{j-1}_i - \alpha \frac{\partial{Q^j}}{\theta_i},
\end{equation}
where $\frac{\partial{Q^j}}{\theta_i}$ is the partial derivative with respect to $\theta_i$, $\theta^j_i$ denotes the update in the jth iteration of $\theta_i$, and $\alpha$ is the learning rate. 
The pseudocode of the proposed method is shown in Algorithm~\ref{alg:nnmethod}.

\begin{algorithm}[h!]
	\caption{Incremental learning for infection rates prediction}
	\label{alg:nnmethod}
	\begin{algorithmic}[1]
            \Statex \textbf{Input:} ``Lookback'' window size $W$, Day ahead to predict $D$
            \State Wait $W$ days to fill window.
            \State Create model $f^W.init()$ \Comment $j = W$
            \State Observe instance $\grave{\beta}^W = \{\beta^W, \beta^{W-1}, ..., \beta^1\}$
            \State Predict $\hat{\beta}^{W+1} = f^W.predict(\grave{\beta}^W)$
    	\For{each day $j \in [W + 1, W + D -1)$} \Comment Only predictions these days
                \State Observe ground truth $\beta^j$
                \State Create instance $\grave{\beta}^j = \{\beta^j, \beta^{j-1}, .., \beta^{j-W+1}\}$
                \State Predict $\hat{\beta}^{j + D} = f^W.predict(\grave{\beta}^j)$ \Comment $f^W$ hasn't been updated yet
		  \EndFor
    	\For{each day $j \in [W + D, \infty)$} \Comment Predictions and training
                \State Observe ground truth $\beta^j$
                \State Incremental training $f^j = f^{j-1}.train((\grave{\beta}^{j-D}, \beta^j))$ \Comment As in Eq (\ref{eq:training})
                \State Create instance $\grave{\beta}^j = \{\beta^j, \beta^{j-1}, .., \beta^{j-W+1}\}$
                \State Predict $\hat{\beta}^{j + D} = f^j.predict(\grave{\beta}^j)$
		  \EndFor
	\end{algorithmic}
\end{algorithm}

\textbf{Evaluation methodology.} All experiments are evaluated according to the widely used regression forecast metrics Mean Absolute Error (MAE) and Mean Absolute Percentage Error (MAPE), as defined in Eqs~(\ref{eq:mae}) and (\ref{eq:mape}) respectively.

\begin{subequations}
\begin{align}
    \label{eq:mae}
    MAE &= \frac{1}{n} \sum_{t = W + D}^T (\hat{\beta}^t - \beta^t), \\
    \label{eq:mape}
    MAPE &= \frac{1}{n} \sum_{t = W + D}^T \frac{|\beta^t - \hat{\beta}^t|}{|\beta^t|} \times 100\%,
\end{align}    
\end{subequations}
\noindent where $n = T - W - D + 1$ is the total number of predictions made.

Furthermore, all experiments are run over 10 repetitions where we present the mean and standard deviation of the relevant metrics. 

\textbf{Data preprocessing.} The infection rates are normalized in the range $[0,1]$ by dividing them with the largest value of the time series. 
Outliers (typically, very small values, e.g., less than 0.1) have been removed through visual inspection of data plots. Linear interpolation is used to fill in missing values (e.g., due to outlier removal) within a window $\grave{\beta}^t$, while the First Observation Carried Backwards and the Last Observation Carried Forwards methods are used for any missing values in the beginning or end of the window, respectively.

\textbf{Learning models.} Each learning model (one for each infection rate) is a Multilayer Perceptron (MLP) model, which is a standard feed-forward fully-connected neural network. Most of the hyper-parameters of the MLPs have the same values, while small differences exist in the rest. All hyper-parameter values are reported in~\nameref{Appendix}.

\textbf{Role of the window size $W$.} We have examined the performance of the proposed method using three window sizes $W \in \{7, 14, 30\}$. In all cases, we forecast each rate $\hat{\beta}^{j+D}$ a week ahead, i.e., $D = 6$. The best performing choice is with a fourteen-day window. The results are presented in  Table \ref{tab:nnperformance}, while the results for all window sizes are shown in \nameref{Appendix}.

\textbf{Performance.} In addition to Table \ref{tab:nnperformance}, the performance of the predicted infection rates compared to the original ones is depicted in Fig~\ref{fig:predictedparameters}. The models appear to perform well, while the fact that some instances are not captured might be attributed to the dynamic, non-stationary nature of the data.

\subsubsection{Infected population prediction and evaluation}
\label{Prediction}
The parameterization of the SIDAREVH model enabled predictions of the vaccinated and unvaccinated infected population. Predictions were made by extrapolating the infection rates identified firstly in section \nameref{EstimationOfParameters} by fitting the model to data, and then using the infection rates predictions through the neural network approach presented in section \nameref{PredictionOfParametersML}. Projections were generated from September 2020 to March 2022, in sliding windows of size $n$, using the information from the corresponding previous windows of size $n$, where $n \in \{7, 14\}$, all moving by one day per increment. The information used for the prediction included the daily vaccinated population and the values of the four infection rates $\beta_{uu}, \beta_{vu}, \beta_{vv}, \beta_{uv}$ that describe the previous $n$-day window. The initial values of the eight states of the model, described by Eq~(\ref{eq:dynamics}), for every predicted sliding window, were the percentage population of Fig~\ref{fig:trajectories}, on the day before every window of prediction.

The SIDAREVH model’s predictive accuracy was validated by the commonly employed metric Mean Absolute Percentage Error (MAPE), which for day $j$ is defined as
\begin{eqnarray}
\label{eq:pe}
	 MAPE_{j}(n,i,d,\hat{i}_{j},\hat{d}_{j})= \frac{1}{n} \sum_{t=j+1}^{j+n}|\frac{[\hat{i}_{j}(t)+\hat{d}_{j}(t)]}{i(t)+d(t)}-1| \\ \times 100\% \nonumber.
\end{eqnarray} 
The MAPE metric is commonly employed in medical research, contributing to the comprehension of the error distributions~\cite{aregay2016multiscale,prates2019spatial,mckenzie2011mean}. $MAPE_{j}$ describes the average percentage error at day $j$, that appears between the data and the predicted infected cases of a time window of size $n$. Terms $i(t)$ and $d(t)$ denote the unvaccinated and vaccinated infected states from the data at time $t$, and terms $\hat{i}_{j}(t)$ and $\hat{d}_{j}(t)$ are the predicted unvaccinated and vaccinated infected states at time $t$ using the knowledge available at time $j$, respectively. The values of $i$ and $d$ are assumed non-negative at all times. Percentage prediction errors  are calculated for every sliding window of the examined period, resulting in a set of percentage errors for all the predicted windows.

\section{Conclusion}
This study developed a novel compartmental model to characterize the progression of pandemics. The proposed SIDAREVH model includes time-varying infection rates, which are the key parameters with the largest impact on the pandemic evolution. We first made estimations of the infection rate values by fitting the available data, regarding the COVID-19 pandemic in Cyprus, to the proposed model and by extrapolating them, we created seven-day infection projections with a mean absolute percentage error of 9.90\%. Then, we used neural regression models trained by incremental learning, to predict the infection rates as they were generated by the model-based estimations, and we observed an improvement in the predictive accuracy of our model, with a mean absolute percentage error of 5.04\%. The resulting high prediction accuracy showcases the reliability of the proposed approach, which is also demonstrated by the small difference between the resulting forecasting error of 5.04\% and the corresponding error that our model can theoretically achieve on the considered dataset, which was estimated to be 3.01\%. Therefore, the proposed hybrid approach, combining compartmental mathematical models and neural networks, consists a reliable forecasting method for the progression of the infected cases. The proposed approach can timely notify  governments and healthcare professionals for a forthcoming crisis, and aid in the design of proper intervention measures that will contain the pandemic's impact, protect the public community and minimize the resulting economic effects.

% or


%\iffalse
%\appendices
\appendix
\label{Appendix}

\section*{Description of symbols used.}
\label{S1_Table_Symols}
\begin{table}[!ht]
    \centering
    \begin{tabular}{|c|l|}
    \hline
         \bf Symbol& \bf Description  \\ \hline
         $s$ & Susceptible state of SIDAREVH model \\ \hline
         $i$ & Infected state of SIDAREVH model \\ \hline
         $d$ & Vaccinated infected state of SIDAREVH model \\ \hline
         $a$ & Hospitalized state of SIDAREVH model \\ \hline
         $r$ & Recovered state of SIDAREVH model \\ \hline
         $e$ & Extinct state of SIDAREVH model \\ \hline
         $v$ & Vaccinated susceptible state of SIDAREVH model \\ \hline
         $h$ & Vaccinated hospitalized state of SIDAREVH model \\ \hline
         $n$ & Model-based window size \\ \hline
         $j$ & Index describing time \\ \hline
         $C_{j}$ & Cost between data and prediction on day $j$ \\ \hline
         $\hat{i}_{j}$ & Predicted infected state on day $j$\\ \hline
         $\hat{d}_{j}$ & Predicted vaccinated infected state on day $j$\\ \hline
         $MAPE_{j}$ & Mean Absolute Percentage Error on day $j$\\ \hline
         $z_{j}$ & Examined window of size $n$\\ \hline
         $k$ & Index describing number of iteration \\ \hline
         $P_{k}$ & Trial parameters set at iteration $k$ \\ \hline
         $\vec{\beta}_{uu}$ & Optimal unvaccinated to unvaccinated infection rate\\ \hline
         $\vec{\beta}_{vu}$ & Optimal vaccinated to unvaccinated infection rate\\ \hline
         $\vec{\beta}_{vv}$ & Optimal vaccinated to vaccinated infection rate\\ \hline
         $\vec{\beta}_{uv}$ & Optimal unvaccinated to vaccinated infection rate\\ \hline
         $W$ & Learning-based window size \\ \hline
         $\grave{\beta}^j$ & Lookback window of size $W$ \\ \hline
         $\beta^{j}$ & Ground truth infection rate at day $j$\\ \hline
         $f_{\theta}$ & A regression model parameterised by ${\theta}$\\ \hline
         $\theta$ & Learning-based model parameter \\ \hline
         $D$ & Days of learning-based prediction of infection rates\\ \hline
         $\hat{\beta}^{j}$ & Predicted infection rate on day $j$\\ \hline
         $Q^j$ & Mean Square Error on day $j$\\ \hline
    \end{tabular}
\end{table}

\section*{The values of the constant parameters involved in the SIDAREVH model.}
\label{S2_Table_ConstantParameters}
\begin{table}[H]
    \centering
    \begin{tabular}{|c|c|l|}
        \hline
        \bf Symbol & \bf Value & \bf Justification\\ \hline
        $\gamma_{\alpha},\gamma_{h}$ & $1/12.4 = 0.081$ & ~\cite{wang2020phase}\\ \hline
        $\gamma_{i},\gamma_{d}$ & $1/14 = 0.071$ & ~\cite{WHOreport}\\ \hline
        $\xi_{i}$ & $0.0053$ & ~\cite{verity2020estimates},~\cite{UNPopulation}\\ \hline
        $\xi_{d}$ & $0.000265$ & ~\cite{vasileiou2021effectiveness},~\cite{hyams2021effectiveness}\\ \hline
        $\mu_{\alpha}$ & $0.0085$ & ~\cite{verity2020estimates}\\ \hline
        $\mu_{h}$ & $0.0085$ & ~\cite{vasileiou2021effectiveness},~\cite{hyams2021effectiveness},~\cite{UKAgency}\\ \hline
    \end{tabular}
\end{table}

\section*{MLP learning parameters for each infection rate $\beta_{uu}$, $\beta_{vu}$, $\beta_{vv}$, $\beta_{uv}$.}
\begin{table}[ht!]
\label{S3_Table_nnmodels}
\centering
        \begin{tabular}{|l|cccc|}
            \hline
             & \multicolumn{1}{c|}{$\beta_{uu}$} & \multicolumn{1}{c|}{$\beta_{vu}$} & \multicolumn{1}{c|}{$\beta_{vv}$} & $\beta_{uv}$ \\ \hline
            \bf Hidden layers & \multicolumn{1}{c|}{[64, 8]} & \multicolumn{1}{c|}{[32, 128, 8]} & \multicolumn{1}{c|}{[128, 64]} & [32, 8, 8] \\ \hline
            \bf Learning Rate & \multicolumn{1}{c|}{0.01} & \multicolumn{1}{c|}{0.01} & \multicolumn{1}{c|}{0.001} & 0.01 \\ \hline
            \bf L2 Regularization & \multicolumn{1}{c|}{0.0001} & \multicolumn{1}{c|}{0.001} & \multicolumn{1}{c|}{0.1} & 0.01 \\ \hline
            \bf Optimizer & \multicolumn{4}{c|}{Adam} \\ \hline
            \bf Weight initialization & \multicolumn{4}{c|}{He Normal} \\ \hline
            \bf Hidden activations & \multicolumn{4}{c|}{Leaky ReLU} \\ \hline
            \bf Output activation & \multicolumn{4}{c|}{ReLU} \\ \hline
            \bf Loss function  & \multicolumn{4}{c|}{MSE} \\ \hline
            \bf Num. of epochs & \multicolumn{4}{c|}{5} \\ \hline
            \bf Mini-batch size & \multicolumn{4}{c|}{1} \\ \hline
        \end{tabular}
\end{table}

\section*{Information about the filtering technique used in the model-based estimation of infection rates involved in SIDAREVH model.}
\label{S1_Appendix}
The infection rates $\beta_{uu}$, $\beta_{vu}$, $\beta_{vv}$, $\beta_{uv}$ included in the SIDAREVH model, and estimated by fitting the model to the available data were filtered by a function of MATLAB, that detects and removes the outliers from data. 
Outliers are defined as elements more than three scaled MAD (Median Absolute Deviation) from the median. The scaled MAD for a vector of parameter values $B$ is defined as $MAD = c \times median(|B-median(B)|),$ where $c = -1/(\sqrt{2} \times erfcinv(3/2))$, where $erfcinv$ is the inverse complementary error function, and the error function is twice the integral of the Gaussian distribution with 0 mean and variance of 1/2.

\section*{MLP performance for each infection rate $\beta_{uu}$, $\beta_{vu}$, $\beta_{vv}$, $\beta_{uv}$.}
Table \ref{Table1_S2} presents the MAPE (S.D.) for each of the four rates, per examined window size.

\begin{table*}[h!]
\renewcommand\thetable{1}
\label{S2_Appendix}
\caption{MAPE for each infection rate per window size (S.D.).}
\centering
    \begin{tabular}{|c|c|c|c|c|}
        \hline
         \bf Window size & \textbf{$\beta_{uu}$} & \textbf{$\beta_{vu}$} & \textbf{$\beta_{vv}$} & \textbf{$\beta_{uv}$} \\ \hline
        \textbf{7} & 6.39\% (7.24\%) & 1.70\% (4.98\%) & 12.54\% (20.05\%) & 10.66\% (12.18\%) \\ \hline
        \textbf{14} & 6.35\% (7.57\%) & 1.48\% (4.05\%) & 15.40\% (17.78\%) & 8.00\% (13.10\%) \\ \hline
        \textbf{30} & 6.40\% (7.17\%) & 1.71\% (5.43\%) & 17.40\% (18.66\%) & 10.35\% (14.93\%) \\ \hline
    \end{tabular}
    \label{Table1_S2}
\end{table*}

\section*{SIDAREVH model's predictive accuracy for fourteen-day windows.}
\label{S3_Appendix}
The error results presented below, concern predictions of the infected population using the model-based infection rates identified at each day of the examined period, using the previous fourteen-day windows data, to the next fourteen-day windows. The model-based estimation made by fitting the available data of Cyprus in the proposed SIDAREVH model. The first row of the tables below refers to the mean and the standard deviation of all predicted sliding windows errors, while the second row excludes the 5\% of the highest error values.

Table~\ref{Table1_S3} presents the ability of the SIDAREVH model to describe the data accurately. Specifically, the error results describe the accuracy of forecasting the infected population, when considering fourteen-day prediction windows, and by using the unfiltered model-based estimated infection rates. These results can be interpreted as the lower bound that the proposed SIDAREVH model may achieve when considering prediction windows of fourteen days on the considered dataset. 

\begin{table}[!ht]
\centering
\renewcommand\thetable{1}
\caption{{\bf SIDAREVH model's accuracy in describing the infected population data using model-based optimized unfiltered time-varying infection rates with fourteen-day windows.}}
    \begin{tabular}{|c|c|c|}
        \hline
         & \bf MAPE & \bf S.D.\\ \hline
        All data & 5.05\% & 3.81\%  \\ \hline
        Lowest 95\% of errors & 4.39\% & 2.52\% \\
        \hline
    \end{tabular}
    \label{Table1_S3}
\end{table}

Table~\ref{Table2_S3} presents the errors of SIDAREVH model's evaluation, while the forecasting procedure used the filtered model-based estimated infection rates with the outliers removed for fourteen-day windows. 

\begin{table}[!ht]
\renewcommand\thetable{2}
\caption{\bf SIDAREVH model’s accuracy in describing the infected population data using model-based optimized filtered time-varying infection rates with fourteen-day windows.}
\centering
    \begin{tabular}{|c|c|c|}
        \hline
         & \bf MAPE & \bf S.D.\\ \hline
        All data & 26.95\% & 23.69\%  \\ \hline
        Lowest 95\% of errors & 23.31\% & 17.98\% \\
        \hline
    \end{tabular}
    \label{Table2_S3}
\end{table}

\section*{Daily MPE of prediction windows.}
\label{S1_Fig}
The MPE values associated with the number of days of prediction windows, when using seven and fourteen days of previous data.
\begin{figure}[H]
    \centering
    \includegraphics[scale=0.5]{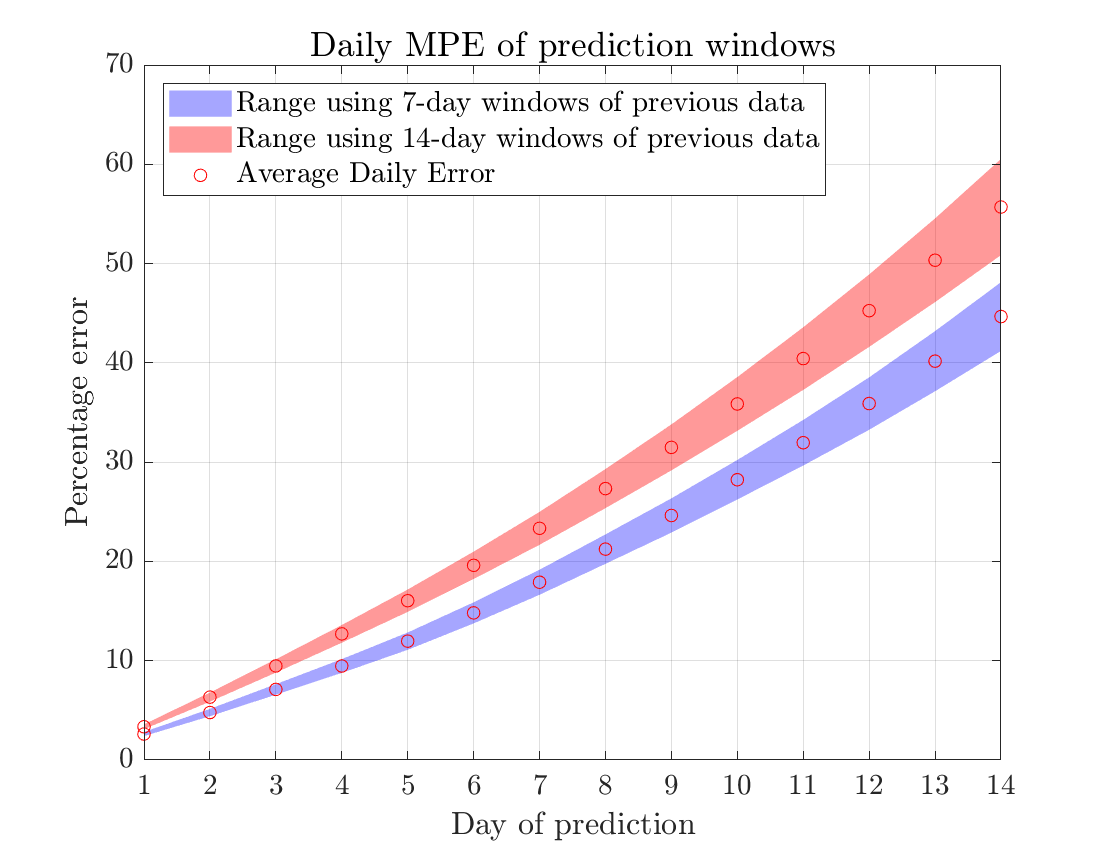}
\end{figure}
%\fi

 \balance
%\section*{Acknowledgment}
%This work was supported by the European Union’s Horizon 2020 research and innovation programme under grant agreement No 739551 (KIOS CoE - TEAMING) and from the Republic of Cyprus through the Deputy Ministry of Research, Innovation and Digital Policy. It was also supported by the CIPHIS (Cyprus Innovative Public Health ICT System) project of the NextGenerationEU programme under the Republic of Cyprus Recovery and Resilience Plan under grant agreement C1.1l2.

 \printbibliography

\ifCLASSOPTIONcaptionsoff
  \newpage
\fi



%\bibliographystyle{IEEEtran}

\end{document}